# Hierarchic Models of Turbulence, Superfluidity and Superconductivity

Alex Kaivarainen

University of Turku, Department of Physics, Turku, Finland

http://www.karelia.ru/~alexk
http://arxiv.org/find/physics/1/au:+Kaivarainen/0/1/0/all/0/1

**Contents**

## Introduction to Hierarchic Theory of Condensed Matter

A basically new hierarchic quantitative theory, general for solids and liquids, has been developed (Kaivarainen, 1995; 2001; 2001a; 2003). It is assumed, that anharmonic oscillations of particles in any condensed matter lead to appearance of three-dimensional (3D) superposition of standing de Broglie waves of molecules, electromagnetic and acoustic waves. Consequently, any condensed matter could be considered as a gas of 3D standing waves of corresponding nature. Our approach unifies and develops strongly the Einstein's and Debye's models.

The most probable de Broglie wave (wave B) length is determined by the ratio of Plank constant to the most probable momentum of molecules or by ratio of its most probable phase velocity to frequency. The waves B in condensed matter are related to molecular translations (tr) and librations (lb). As far corresponding oscillations of molecules in condensed matter are anharmonic, they do not follow classical Maxwell-Boltzmann distribution and the most probable de Broglie wave length can exceed the distance between centers of molecules many times. *This makes possible the atomic and molecular mesoscopic Bose condensation (mBC) in regular solids and liquids at temperatures, below boiling point. It is one of the most important results, which is discovered by computer simulations, based on our theory, for examples of water and ice.*

Four strongly interrelated new types of quasiparticles (collective excitations) were introduced in our hierarchic model:

1. *Effectons (tr and lb)*, existing in "acoustic" (a) and "optic" (b) states represent the coherent clusters in general case**;**

2. *Convertons*, corresponding to interconversions between *tr* and *lb* types of the effectons (flickering clusters);

3. *Transitons* are the intermediate $[a \rightleftharpoons b]$ transition states of the *tr* and *lb* effectons;

4. *Deformons* are the 3D superposition of IR electromagnetic or acoustic waves, activated by *transitons* and *convertons*.

*Primary effectons* (tr and lb) are formed by 3D superposition of the most probable standing de Broglie waves of the oscillating ions, atoms or molecules. The volume of effectons (tr and lb) may contain from less than one, to tens and even thousands of molecules. The first condition means validity of classical approximation in description of the subsystems of the effectons. The second one points to quantum properties of coherent clusters due to molecular Bose condensation on mesoscopic spatial scale.

The liquids are semiclassical systems because their primary (tr) effectons contain less than one molecule and primary (lb) effectons - more than one molecule. The solids are quantum systems totally because both kind of their primary effectons (tr and lb) are molecular Bose condensates. These consequences of our theory are confirmed by computer calculations.

The 1st order $[gas \rightarrow liquid]$ transition is accompanied by strong decreasing of rotational (librational) degrees of freedom due to origination of primary librational effectons and $[liquid \rightarrow solid]$ transition - by decreasing of translational degrees of freedom due to Bose-condensation of primary (tr) effectons.

In the general case the effecton can be approximated by parallelepiped with edges corresponding to de Broglie waves length in three selected directions (1, 2, 3), related to the symmetry of the molecular dynamics. In the case of isotropic molecular motion the effectons' shape may be approximated by cube.

The edge-length of primary effectons (tr and lb) or number of molecules in its volume can be considered as the "parameter of order".

The in-phase oscillations of molecules in the effectons correspond to the effecton's (a) - acoustic state. The counterphase molecular oscillations correspond to effectons (b) - optic state. States (a) and (b) of the effectons differ in potential energy only, however, their kinetic energies, momentums and spatial dimensions - are the same. The b-state of the effectons has a common feature with Frölich's polar mode.

The $(a \rightarrow b)$ or $(b \rightarrow a)$ transition states of the primary effectons (tr and lb), defined as primary transitons, are accompanied by a change in molecule polarizability and dipole moment without density fluctuations. At this case they lead to absorption or radiation of IR photons, respectively.

Superposition of three internal standing IR photons of different directions (1,2,3) - forms primary electromagnetic deformons (tr and lb). On the other hand, the [lb$\rightleftharpoons$ tr] convertons and secondary transitons are accompanied by the density fluctuations, leading to absorption or radiation of phonons.

Superposition resulting from interception of standing phonons in three directions (1,2,3), forms secondary acoustic deformons (tr and lb).

*Correlated collective excitations* of primary and secondary effectons and deformons (tr and lb)**,** localized in the volume of primary *tr* and *lb electromagnetic* deformons**,** lead to origination of *macroeffectons, macrotransitons and macrodeformons* (tr and lb respectively)**.**

*Correlated simultaneous excitations of tr and lb macroeffectons* in the volume of superimposed *tr* and *lb* electromagnetic deformons lead to origination of *supereffectons.*

In turn, the coherent excitation of *both: tr* and *lb macrodeformons and macroconvertons* in the same volume means creation of *superdeformons.* Superdeformons are the biggest (cavitational) fluctuations, leading to microbubbles in liquids and to local defects in solids.

Total number of quasiparticles of condensed matter equal to 4!=24, reflects all of possible combinations of the four basic ones [1-4], introduced above. This set of collective excitations in the form of "gas" of 3D standing waves of three types: de Broglie, acoustic and electromagnetic - is shown to be able to explain virtually all the properties of all condensed matter.

The important positive feature of our hierarchic model of matter is that it does not need the semiempirical intermolecular potentials for calculations, which are unavoidable in existing theories of many body systems. The potential energy of intermolecular interaction is involved indirectly in dimensions and stability of quasiparticles, introduced in our model.

The main formulae of theory are the same for liquids and solids and include following experimental parameters, which take into account their different properties:

[1]- Positions of (tr) and (lb) bands in oscillatory spectra;

[2]- Sound velocity;

[3]- Density;

[4]- Refraction index (extrapolated to the infinitive wave length of photon).

The knowledge of these four basic parameters at the same temperature and pressure makes it possible using our computer program, to evaluate more than 300 important characteristics of any condensed matter. Among them are such as: total internal energy, kinetic and potential energies, heat-capacity and thermal conductivity, surface tension, vapor pressure, viscosity, coefficient of self-diffusion, osmotic pressure, solvent activity, etc. Most of calculated parameters are hidden, i.e. inaccessible to direct experimental measurement.

The new versions of Brillouin light scattering and Mössbauer effect are created on the basis of hierarchic theory (Kaivarainen, 1995; 2001).

Some original aspects of water in organization and large-scale dynamics of biosystems - such as proteins, DNA, microtubules, membranes and regulative role of water in cytoplasm, cancer development, quantum neurodynamics, etc. have been analyzed in the framework of Hierarchic theory (Kaivarainen, 2003; 2003a).

Computerized verification of our Hierarchic concept of matter on examples of water and ice is performed, using special computer program: Comprehensive Analyzer of Matter Properties (CAMP, copyright, 1997, Kaivarainen). The new optoacoustic device (CAMP), based on this program, with possibilities much wider, than that of IR, Raman and Brillouin spectrometers, has been proposed (Kaivarainen, 2003; 2004).

This is the first theory able to predict all known experimental temperature anomalies for water and ice. The quantitative conformity between theory based computer calculations and experiment for lot of physical parameters is very good without adjustable parameters. This represents in fact strong experimental evidence in proof of our theory. A series of paper, related to Hierarchic theory is on-line: http://arXiv.org/find/physics/1/au:+Kaivarainen/0/1/0/all/0/1

The hierachical scenarios of turbulence, superconductivity and superfluidity have been elaborated (Kaivarainen, 1995). Their revised form are presented in this paper.

## 1. Turbulence. General description

The type of flow when particles move along the straight trajectory without mixing with adjoining layers, is termed laminar flow.

If the layers of the liquid of the laminar flow are moving relative to each other at different velocities, then the forces of internal friction ($F_{fr}$) or viscosity forces originate between them:

$$F_{fr} = \eta \left| \frac{\Delta \mathbf{v}}{\Delta \mathrm{d}} \right| S, \qquad\qquad 1$$

where: $\Delta \mathbf{v}$ is relative liquid layer velocity; $S$ is the contact surface; $\eta$ is dynamic viscosity; $\left| \frac{\Delta v}{\Delta d} \right|$ is the module of velocity gradient directed to the surface of layers.

Near the walls of a straight tube the velocity of laminar flow is equal to zero and in the center of the tube it is maximum.

The relation between the layer velocity and its distance from central axes of the tube (r) is parabolic:

$$\mathbf{v}(r) = \mathbf{v}_0 \left( 1 - \frac{r^2}{a_t^2} \right) \qquad\qquad 2$$

where: $a_t$ is tube radius; $\mathbf{v}_0$ is the velocity of the liquid on the central axis, depending on the difference of pressure at the ends of the tube:

$$\Delta P = P_1 - P_2 \qquad\qquad 3$$

as follows:

$$\mathbf{v}_0 = \frac{P_1 - P_2}{4\eta l} a_t^2 \qquad\qquad 4$$

where: $l$ is tube length and $\eta$ is dynamic viscosity.

The flux of liquid, i.e. the volume of liquid flowing over the cross-section of the tube during a time unit is determined by the Poiseuille formula:

$$Q = \frac{(P_1 - P_2)\pi a_t^2}{8\eta l} \qquad\qquad 5$$

This formula has been used frequently for estimation of dynamic viscosity $\eta$.

The corresponding mass of flowing liquid is equal:

$$m = \rho Q \qquad\qquad 6$$

and corresponding kinetic energy:

$$T_k = \frac{\rho}{4} Q v_0^2 \qquad\qquad 7$$

where: $\rho$ is density of liquid.

The work of internal friction force is:

$$A = -4\eta v_0 l Q / \rho R^2$$

In the case of the laminar movement of a spherical body relative to liquid the force of internal friction (viscosity force) is determined by the Stokes law:

$$F_{fr} = 6\pi a v \eta, \qquad\qquad 8$$

where: (a) is the radius of sphere and $\mathbf{v}$ is its relative velocity.

As a result of liquid velocity ($\mathbf{v}$) and/or the characteristic dimension (a) increasing, the laminar type of liquid flow could change to the turbulent one.

This begins at certain values of the dimensionless Reynolds number:

$$R = \rho \mathbf{v}_c a / \eta = \mathbf{v}_c a / \nu, \qquad\qquad 9$$

where: $\rho$ is liquid density; $\mathbf{v}_c$ is characteristic (average) flow velocity; $\nu = \eta/\rho$ is **kinematic viscosity** of liquid.

For a round tube with radius (a) the critical value of R is about 1000. A turbulent type of flow is accompanied by rapid irregular pulsations of liquid velocity and pressure, representing a kind of self-organization.

In the case of nonstationary movement, the flow can be characterized by two additional dimensionless parameters like:

*Strouhal number:*

$$S = \mathbf{v}_c \tau / a \qquad\qquad 10$$

where: $\tau$ is the characteristic time of velocity ($\mathbf{v}_c$) pulsations;
and *March number*:

$$M = \mathbf{v}_c/\mathbf{v}_s \qquad\qquad 11$$

where: $\mathbf{v}_s$ is sound velocity in liquid.

## 2. Hierarchic mechanism of turbulence

The physical scenario of transition from a laminar flow to a turbulent one is still unclear. It is possible to propose the mechanism of this transition based on hierarchic concept of matter.

Let us start from the assumption that in the case of laminar flow, the thickness of parallel layers is determined by the thickness of primary electromagnetic deformons (translational $\sim 5 \cdot 10^5 \mathring{A}$ and librational $\sim 10^5 \mathring{A}$), equal to linear dimensions of corresponding macrodeformons (Kaivarainen, 1995; 2001).

The total internal energy and the internal pressure of neighboring layers are not equal.

The surface between such layers can be characterized by corresponding surface tension $[\sigma(r)]$. Surface tension prevents mixing between layers with different laminar flow velocities.

According to our model, the thickness of the two outer borders of each layer (skin-surface) is determined by the effective linear dimensions of primary librational effectons $[l_{tr,lb} \sim 15\mathring{A}\ at\ 25^0C]$, related to the corresponding most probable wave B length ($\lambda_{tr,lb}$) of molecules in liquid:

$$l_{tr,lb} = (V_{ef}/V_{ef}^{2/3}) \qquad\qquad 12$$

Decreasing of $l_{lb}$, depending on most probable momentum of liquid molecules, as a result of increased flow velocity $[\mathbf{v}_1(r)]$ see eq. (13) and/or temperature elevation, in accordance with our theory of surface tension (Kaivarainen, 2001; http://arxiv.org/abs/physics/0102086), leads to reducing of surface tension $\sigma(r)$. In turn, this effect strongly decreases the work of cavitational fluctuations, i.e. the bubbles formation (27) and increases their concentration (29). *These bubbles lead to mixing of laminar layers, the instability of laminar flow and its turning to the turbulent type.*

The critical flow velocity: $\mathbf{v}_c = \mathbf{v}^1(\mathbf{r})$ is determined by the critical *librational* wave B length:

$$\lambda_{lb}^{1,2,3} = h/\left\{ m\left[ (\mathbf{v}_{gr}^{1,2,3})_{lb} + \mathbf{v}^1(r) \right] \right\} \qquad\qquad 13$$

corresponding to the condition of liquid-gas first order phase transition (Kaivarainen, 1995; 2001):

$$[(V_{ef})_{\text{lib}}/(V_0/N_0)] = \left[ \frac{9}{4\pi} (\lambda^{(1)} \cdot \lambda^{(2)} \cdot \lambda^{(3)})_{\text{lib}}/(V_0/N_0) \right] \leq 1 \qquad\qquad 14$$

where: $m$ is molecular mass; $\mathbf{v}_{gr}^{1,2,3}$ is the most probable librational group velocity of molecules of liquid in selected directions $(1,2,3)$; $\mathbf{v}^1(r)$ is the flow velocity of a liquid layer in the tube at the distance (r) from the central axes of the tube (2).

Increasing of $\mathbf{v}^1(r)$ at $r \to 0$ decreases $\lambda_{lb}^1$ and $(V_{ef})_{lb}$ in accordance with (13) and (14). The value of $\lambda_{lb}^1$ is also related to phase velocity ($\mathbf{v}_{ph}^a$) and frequency ($\nu_1^a$) of the primary librational effecton in (a) state:

$$\lambda_{\text{lib}}^1 = h/\{m[(\mathbf{v}_{gr}^1)_{lb} + \mathbf{v}^1(r)]\} = \left( \frac{\mathbf{v}_{\text{ph}}^a}{\nu_1^a} \right)_{lb} = (\mathbf{v}_{ph}^a/\nu_p^1)\left[ \exp(h(\nu_p^1)_{lb}/kT) - 1 \right] \qquad\qquad 15$$

where:

$$(\nu_p^1)_{lb} = (\nu_1^b - \nu_1^a)_{lb} = c(\nu_p^1)_{lb} \qquad\qquad 16$$

is the frequency of $(a \Leftrightarrow b)_{lb}$ transitions of the primary librational effecton of a flowing liquid determined by librational band wave number $(\tilde{\nu}_p^1)_{lb}$ in the oscillatory spectra.

It was calculated earlier for water under stationary conditions, that the elevation of temperature from $0^0$ to $100^0 C$ till the phase transition condition (14), is accompanied by the increase in $(\mathbf{v}_{gr})_{lb}$ from $(1.1$ to $4.6) \cdot 10^3 cm/s$ (Kaivarainen, 1995; 2001).

This means that at $30^0 C$, when $(\mathbf{v}_{gr})_{lb} \simeq 2 \cdot 10^3 cm/s$, the critical flow velocity $\mathbf{v}^1(r)$, necessary for mechanical boiling of water (condition 14), should be about $2.5 m/s$.

The reduced number of primary librational effectons $(N_{ef})$ in the volume $(V_D^M)$ of primary electromagnetic deformons (tr and lib) also increases with temperature and/or flow velocity:

$$\left( N_{ef} \right)_{tr,lb}^D = \left[ \frac{P_a + P_b}{Z} n_{ef} \cdot V_D^M \right] \qquad 17$$

The reduced number of primary transitons $(N_t)$ has a similar dependence on T and $\mathbf{v}^1(r)$, due to increasing of $n_{ef}$, and $V_D^M$ as far:

$$\left( N_t \right)_{tr,lb}^D = \left( N_{ef} \right)_{tr,lb}^D \qquad 18$$

The analysis of eq.(15) *predicts* that at permanent temperature (T = *const)* an increase in $\mathbf{v}^1(r)$ must be accompanied by the low-frequency shift of the librational band: $\tilde{\nu}_{lib}^{(1)} \simeq 700 cm^{-1}$ and/or by the decrease in sound velocity $(\mathbf{v}_{ph}^a)_{lb}$ in the direction of flow. This consequences of model can be verified experimentally.

It follows also from our hierarchic theory that these changes should be accompanied by a rise in dynamic viscosity due to increased structural factor $(\mathbf{T}_{kin}/\mathbf{U}_{tot})$.

Turbulent pulsations of flow velocity $(\Delta \mathbf{v})$ originate under developed turbulence conditions:

$$\Delta \mathbf{v} = \mathbf{v}_{tur} = \mathbf{v} - \overline{\mathbf{v}} \qquad 19$$

where: $\overline{\mathbf{v}}$ is average flow velocity and $(\mathbf{v})$ is instant flow velocity.

The frequencies of large-scale pulsations have the order of:

$$\nu = \overline{\mathbf{v}}/\lambda_p, \qquad 20$$

where: $\lambda_p$ is the main scale of pulsations.

$\lambda_p$ can correlate with the dimensions of electromagnetic deformons, introduced in Hierarchic theory (Kaivarainen, 2001) and can be determined by the transverse convection rate, depending on the bubbles dimensions.

*The pulsations of flow velocity ($\Delta \mathbf{v}$) can result from:*
*a) mixing of parallel layers with different flow velocity and*
*b) fluctuation of viscosity force (eq. 8) due to fluctuations in bubbles radius and concentration, as well as density, viscosity and thermal conductivity;*
*c) movements and appearance of the bubbles as a result of the Archimedes force.*

The bubbles have two opposite types of influence on *instant* velocity. The layer mixing effect induced by them can increase flow velocity.

On the other hand, the bubbles can simultaneously decrease flow velocity due to enhanced internal friction.

In the case of developed turbulence with different scales of pulsations it is reasonable to introduce the characteristic Reynolds number:

$$R_\lambda = \mathbf{v}_{tur}\lambda_p/\nu_{tur} \qquad 21$$

where: $\mathbf{v}_{tur}$ is the velocity of pulsation and $\nu_{tur} = (\eta/\rho)_{tur}$ characteristic kinematic viscosity in turbulence conditions.

The ratio between turbulent kinematic viscosity $(\nu_{tur})$ and a laminar one $(\nu)$ is related to the corresponding Reynolds numbers like (Landau, Lifshits, 1988):

$$\frac{v_{tur}}{v} \sim \frac{R}{R_{tur}} \qquad\qquad 22$$

Large-scale pulsations correspond to high $\lambda_p$ values and low $v_{tur}$ values. i.e. high characteristic turbulence Reynolds numbers (see 21).

According to our model, the maximum energy dissipation occurs in the volume of superdeformons.

*Mechanically induced boiling* under conditions of turbulence (eqs. 13 and 14) is accompanied by the appearance of gas bubbles related to the increased superdeformons probability and decreased surface tension between layers.

Critical bubble creation work ($W$) is strongly dependent on inter- layer surface tension ($\sigma$). A general classical theory (Nesis, 1973) gives:

$$W = \frac{4}{3}\pi a^2 \sigma = \frac{16\pi\sigma^3}{3(P - P_{ext})^2} \qquad\qquad 27$$

where:

$$P = P_{ext} + \frac{2\sigma}{a} = P_{a=\infty} \cdot \exp\left(-\frac{2\sigma V_e}{a\,kT}\right) \qquad\qquad 28$$

P is the internal gas pressure in a bubble with radius ($a$); $V_e$ is the volume of liquid occupied by one molecule.

The bubbles quantity ($N_b$) has an exponential dependence on $W$:

$$N_b = \exp\left(-\frac{W}{kT}\right) \qquad\qquad 29$$

One can see from our hierarchic theory of surface tension (Kaivarainen, 2001) that under *mechanical boiling conditions* the skin-surface thickness (12) : $l \to (V_0/N_0)^{1/3}$ and $q^s \to 1$, the interlayer surface tension ($\sigma$) tends to zero, $W$ decreases and $N_b$ increases.

We can conclude that the hierarchic scenario of mechanical boiling, presented here can provide a background for elaboration of a quantitative physical theory of turbulence and other hydrodynamic instabilities like Taylor's and Benar's ones.

### 3. Superfluidity. General description

Superfluidity has been revealed for two liquids only: helium isotopes: $^4\textbf{He}$ with boson's properties ($S = 0$) and $^3\textbf{He}$ with fermion properties ($S = 1/2$). The interactions between the atoms of these liquids is very weak. It will be shown below that the values of normal sound velocity at temperatures higher than those of second order phase transition ($\lambda$-point) are lower than the most probable thermal velocities of the atoms of these liquids.

The first theories of superfluidity were proposed by Landau (1941) and Feynman (1953).

First order phase transition [gas → liquid] occurs at $4.22 K$.

Second order phase transition, when superfluidity originates, $^4\textbf{He} \to \textbf{HeII}$ takes place at $T_\lambda = 2.17 K$ ($P_{ext} = 1$ atm.). This transition is accompanied by:

a) heat capacity jump to higher values;

b) abruptly increased thermal conductivity;

c) markedly decreased cavitational fluctuations and bubbles in liquid helium.

For explanation of experimental data Landau supposed that at $T < T_\lambda$ the He II consists of two components:

- the *superfluidity component* with relative fraction of density $\rho_S/\rho$, increasing from zero at $T = T_\lambda$ to 1 at T = 0 K. The properties of this component are close to those of an ideal liquid with a potential type of flow. The entropy of this component is zero and it does not manifest the viscous friction on flowing through narrow capillaries;

- the *normal component* with density

$$\rho_n = \rho - \rho_s \qquad\qquad 30$$

decreasing from 1 at $T = T_\lambda$ to zero at T = 0 K. This component behaves as a usual viscous liquid which exhibits dumping of the oscillating disk in He II. Landau considered this component to be a gas of two types of excitations: *phonons* and *rotons*.

The hydrodynamics of normal and superfluid components of He II are characterized by *two velocities*: normal ($\mathbf{v}_n$) and superfluid one:

$$\mathbf{v}_{sf} = (\hbar/m)\nabla\varphi \qquad\qquad 31$$

where $\nabla\varphi \sim k_{sf} = 1/L_{sf}$ is a phase of Bose-condensate wave function - see eq. 36.

As a result of two types of hydrodynamic velocities and densities, the corresponding 2 types of sound waves propagate in the volume of **He II.**

The *first sound* ($U_1$) is determined by the usual formula valid for normal condensed matter:

$$U_1^2 = (\partial P/\partial\rho)_S \qquad\qquad 32$$

In this case density oscillations spread in the form of phonons.

The *second sound* ($U_2$) is related to oscillations of temperature and entropy (S):

$$U_2^2 = \rho_S TS^2/c\rho_n \qquad\qquad 33$$

In normal condensed media the temperature oscillation fade at the distance of the order of wave length.

Landau considered the second sound as density waves in the gas of quasiparticles: *rotons and phonons.*

The third sound ($U_3$) propagates in the thin surface films of He II in the form of "ripplons", i.e. quantum capillary waves related to the isothermal oscillations of the superfluid component.

$$U_3 = (\rho_S/\rho_S) \cdot d \cdot \frac{\partial \mathrm{E}}{\partial \mathrm{d}}(1 + TS/L), \qquad\qquad 34$$

where: $(\bar\rho_S/\rho_S)$ is the relative density of superfluid component averaged in the thickness of the film (d); E is the potential of Van- der-Waals interactions of $^4$He atoms with the bottom surface; L is evaporation heat.

The *fourth sound* ($U_4$) propagates in He II, located in very narrow capillaries, when the length of quasiparticles (phonons and rotons) free run is compatible or bigger than the diameter of these capillaries or pores.

The hydrodynamic velocity ($\mathbf{v}_n$) of the normal component under such conditions is zero and $\rho_n/\rho \ll \rho_{sf}/\rho$:

$$U_4^2 = (\rho_S/\rho)U_1^2 + (\rho_n/\rho)U_2^2 \simeq (\rho_S/\rho)U_1^2 \qquad\qquad 35$$

In accordance with Bose-Einstein statistics, a decrease in temperature, when $T \to T_\lambda$, leads to condensation of bosons in a minimum energy state.

This process results in the origination of a superfluid component of He II with the coherent thermal and hydrodynamic movement of atoms.

Coherence means that this movement can be described by the single wave function:

$$\psi = \rho_S^{1/2} \cdot e^{i\varphi} \qquad\qquad 36$$

The movement of the superfluid component is *potential* as far its velocity ($\vec{\mathbf{v}}_{sf}$) is determined by eq.31 and:

$$\mathrm{rot}\ \mathbf{v}_{sf} = 0 \qquad\qquad 37$$

### Vortex filaments in He II

*When the rotation velocity of a cylindrical vessel containing He II is high enough, then the emergency of so-called vortex filaments becomes thermodynamically favorable. The filament is formed by the superfluid component of He II in such a way that their momentum of movement decreases the total energy of He II in a rotating vessel.* The shape of filaments in this case is like a straight rod and their *thickness* is of the order of atom's dimensions, increasing with lowering the temperature at $T < T_\lambda$.

Vortex filaments are continuous. They are closed or limited within the boundaries of a liquid. For each surface surrounding a vortex filament the condition (37) is valid.

The values of velocity of circulation around the axis of filaments are determined (Landau, 1941) as follows:

$$\oint \mathbf{v}_{sf} dl = 2\pi r \mathbf{v}_{sf} = 2\pi \kappa \qquad\qquad 38$$

and

$$\mathbf{v}_{sf} = \kappa / r \qquad\qquad 39$$

Increasing the radius of circulation (r) leads to decreased circulation velocity ($\mathbf{v}_{sf}$). Substituting $\mathbf{v}_{sf}$ in eq.31, we obtain:

$$\oint \mathbf{v}_{sf} dl = \frac{\hbar}{m} \Delta\Phi, \qquad\qquad 40$$

where: $\Delta\Phi = n2\pi$ is a phase change as a result of circulation, $n = 1, 2, 3 \ldots$ is the integer number.

Comparing (40) and (38) gives:

$$\kappa = n \frac{\hbar}{m} \qquad\qquad 41$$

It has been shown that only curls with $n = 1$ are thermodynamically stable.

Taking this into account, we have from (39) and (41):

$$r = n \frac{\hbar}{m \mathbf{v}_{sf}} \qquad\qquad 42$$

An increase in the angle frequency of rotation of the cylinder containing **HeII** results in the increased density distribution of vortex filaments on the cross-section of the cylinder.

As a result of interaction between the filament and the normal component of **HeII**, the filaments move in the rotating cylinder with normal liquid.

The flow of **He II** through the capillaries can be accompanied by appearance of vortex filaments.

In ring-shaped vessels the circulation of closed vortex filaments is stable. Stability is related to the quantum pattern of circulation change (eqs. 38 and 41).

Let us consider now the phenomena of superfluidity in **He II** in the framework of our hierarchic concept.

### 4. Hierarchic scenario of superfluidity

It will be shown below how our hierarchic model (Table 1) can be used to explain He II properties, its excitation spectrum (Fig. 1), increased heat capacity at $\lambda$-point and the vortex filaments formation.

We assume here, that the formulae obtained earlier (Kaivarainen, 2001) for internal energy ($U_{tot} - eq.4.3$), viscosity, thermal conductivity and vapor pressure remain valid for both components of He II.

The theory proposed by Landau (Lifshits, Pitaevsky, 1978) qualitatively explains only the lower branch (a) in the spectrum (Fig. 1), as a result of phonons and rotons excitation.

But the upper branch (b) points that the real process is more complicated and needs introduction of other quasiparticles and excited states for its explanation.

Our hierarchic model of superfluidity interrelates the lower branch with the ground acoustic (a) state of primary effectons in liquid $^4$He and the upper branch with their excited optical (b) state. In accordance with our model, the dissipation and viscosity friction (see section 11.6) arise in the normal component of He II due to thermal phonons radiated and absorbed in the course of the $\bar{b} \to \bar{a}$ and $\bar{a} \to \bar{b}$ transitions of secondary effectons, correspondingly.

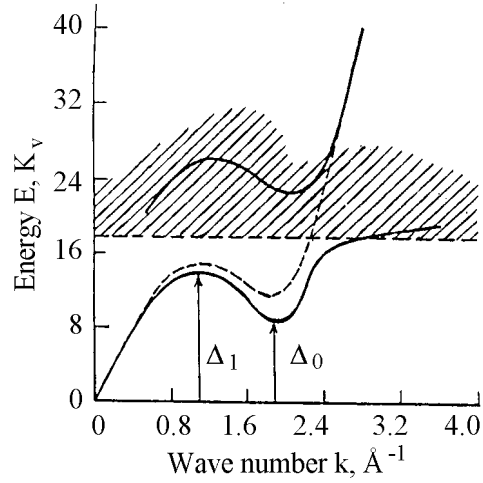

**Fig. 1.** Excitation spectrum of liquid $^4$**He** from neutron scattering measurements (March and Parrinello, 1982). Spectrum is characterized by two branches, corresponding to (a) and (b) states of the primary effectons according to the hierarchic model.

Landau described the minimum in the region of $\lambda$-point using the expression:

$$E = \Delta_0 + \frac{(P - P_0)^2}{2m^*}, \qquad 43$$

where $\Delta_0$ and $P_0$ are the energy and momentum of liquid $^4$He at $\lambda$-point (Fig. 1) and $m^* = 0.16m$ is the effective mass of the $^4$**He** atom ($m_{He} = 4 \cdot 1.44 \cdot 10^{-24}g = 5.76 \cdot 10^{-24}g$). The effective mass $m^*$ can be determined experimentally.

Feynman (1953) explained the same part of the excitation spectra by the non-monotonic behavior of the structure factor $S(k)$ and the formula:

$$E = \hbar\omega = \frac{\hbar^2 k^2}{2mS} = \frac{\hbar^2}{2mL^2 S} \qquad 44$$

where:

$$k = 1/L = 2\pi/\lambda \qquad 45$$

is the wave number of neutron interacting with liquid $^4$He.

Our hierarchic model allows to unify Landau's and Feynman's approaches. The total energy of de Broglie wave either free or as part of condensed matter can be expressed through its amplitude squared ($A^2$), length squared ($L^2$) and effective mass ($m^*$) in the following manner (Kaivarainen, 2001):

$$E_{tot} = T_k + V = m^* \mathbf{v}_{gr} \mathbf{v}_{ph} = \frac{\hbar^2}{2mA^2} = \frac{\hbar^2}{2m^* L^2} \qquad 46$$

where $\mathbf{v}_{gr}$ and $\mathbf{v}_{ph}$ are the most probable group and phase velocities of de Broglie wave.

In accordance with our Hierarchic theory (Kaivarainen, 2001), the structural factor S(k) is equal to the kinetic ($T_k$) to total ($E_{tot}$) energy ratio of wave B:

$$S = T_k/E_{tot} = A^2/L^2 = m^*/m \qquad 47$$

where:

$$T_k = P^2/2m = \frac{\hbar}{2mL} \qquad 48$$

Combining (46), (47) and (48), we obtain the following set of equation for the energy of $^4$He at transition $\lambda$-point:

$$\left.\begin{array}{c} \Delta_0 = E_0 = \frac{\hbar^2}{2mA_0^2} = \frac{\hbar^2}{2m^*L_0^2} \\[2mm] \Delta_0 = \frac{\hbar^2}{2mL_0^2S} = \frac{T_k^0}{S} \end{array}\right\} \qquad 49$$

These approximate formulae for the total energy of liquid $^4$He made it possible to estimate the most probable wave B length, forming the primary librational (or rotational effectons) at $\lambda$-point:

$$\lambda_0 = \frac{h}{mv_{gr}^0} = 2\pi L_0 = 2\pi A_0 \Big( m/m^* \Big)^{1/2}, \qquad 50$$

where the critical amplitude of wave B:

$$A_0 = \hbar \Big( \frac{1}{2mE_0} \Big)^{1/2} \qquad 51$$

can be calculated from the experimental $E_0$ values (Fig.1). Putting in (51) and (50) the available data:

$$\Delta_0 = E_0 = k_B \cdot 8.7K = 1.2 \cdot 10^{-15} \text{ erg};$$

the mass of atom: $m(^4\text{He}) = 5.76 \cdot 10^{-24}g$ and $(m^*/m) = 0.16$, we obtain:

$$\lambda_0 \cong 14 \cdot 10^{-8} cm = 14 \text{Å} \qquad 52$$

the corresponding most probable group velocity of $^4$**He** atoms is: $v_{gr}^0 = 8.16 \cdot 10^3 cm$/s.

It is known from the experiment that the volume occupied by *one atom of liquid* $^4$**He** is equal: $V_{^4\text{He}} = 46\text{Å}^3$/atom. The edge length of the corresponding cubic volume is:

$$\mathbf{l} = \Big( V_{^4He} \Big)^{1/3} = 3.58\text{Å} \qquad 53$$

From (52) and (53) we can calculate the number of $^4$**He** atoms in the volume of primary librational (rotational) effecton at $\lambda$-point:

$$n_V^0 = \frac{V_{ef}}{V_{^4\text{He}}} = \frac{(9/4\pi)\lambda_0}{l^3} = 43 \text{ atoms} \qquad 54$$

One edge of such an effecton contains: $q = (43)^{1/3} \cong 3.5$ atoms of liquid $^4$**He**.

We must take into account, that these parameters can be *lower than the real* ones, as far in above simple calculations we did not consider the contribution of secondary effectons, transitons and deformons to total internal energy.

On the other hand, in accordance with Hierarchic model, the conditions of the maximum stability of primary effectons correspond to the *integer* number of particles in the edge of these effectons (Kaivarainen, 2001).

Consequently, we have to assume that the true number of $^4$*He* atoms forming a primary effecton at $\lambda$-point is equal to $n_V^0 = 64$. It means that the edge of cube as the effecton shape approximation contains $q^0 = 4$ atoms of $^4$*He*:

$$n_e^0 = (n_V^0)^{1/3} = 64^{1/3} = 4 \qquad\qquad 56$$

The primary librational effectons of such a type may correspond to rotons introduced by Landau to explain the high heat capacity of He II.

The thermal momentums of $^4$He atoms in these coherent clusters can totally compensate each other and the resulting momentum of primary effectons is equal to zero. Further decline in temperature gives rise to dimensions of primary effectons, representing *mesoscopic Bose condensate* (mBC). The most stable of them contain in their ribs the integer number of helium atoms:

$$q = q^0 + n \qquad\qquad 56a$$

where: $n = 1, 2, 3\ldots$

$\lambda_0$, $n_V^0$ and $n_e^0$ can be calculated more accurately, using our computer program, based on Hierarchic theory, if the required experimental data on IR spectroscopy and sound velocimetry are available.

### 5. Superfluidity, as a hierarchical self-organization process

Let us consider now the consequence of the phenomena observed in $^4$He in the course of temperature decline to explain Fig. 1 in the framework of hierarchic model:

**1.** In accordance with our model lowering the temperature till the 4.2 K and gas-liquid first order phase transition occurs under condition (6.6). This condition means that the most probable wave B length of atoms related to their rotations or librations starts to exceed the average distance between $^4$He atoms in a liquid phase and mesoscopic Bose condensation (mBC) in form of coherent atomic clusters becomes possible:

$$\lambda = h/mv_{gr} \geq 3.58 \text{Å} \qquad\qquad 57$$

The corresponding value of the most probable group velocity is

$$\mathbf{v}_{gr} \leq 3.2 \cdot 10^4 cm/\text{s}.$$

The translational thermal momentums of particles are usually bigger and waves B length smaller than those related to librations. In accordance with our model of first order phase transitions (Section 6.2 of [1]), this fact determines the difference in the temperatures of [gas → liquid] and [liquid → solid] transitions.

The freezing of liquid $^4$He occurs at a sufficiently high pressure of $\sim 25$ atm only and means the emergency of primary translational effectons (Kaivarainen, 2001). The pressure increasing, as well as temperature decreasing, decline thermal momentum of particles and stimulates Bose condensation, responsible for coherent clusters formation.

In normal component of liquid $^4$**HeII**, like in a usual liquid at $T > 0 \, K$, the existence of primary and secondary effectons, convertons, transitons and deformons is possible. The contributions of each of these quasiparticles determine the total internal energy, kinetic and potential energies, viscosity, thermal conductivity, vapor pressure and many other parameters (Kaivarainen, 2001).

We assume that the lower branch in the excitation spectrum of Fig. 1 reflects the acoustic (a) state and the upper branch the optic (b) state of primary (lb and tr) effectons.

**2.** Decreasing the temperature to $\lambda$-point: $T_\lambda = 2.17K$ is accompanied by condition (55), which stimulates Bose-condensation of atoms, increasing the dimensions of primary effectons. This leads to emergency of primary polyeffectons as superfluid subsystem due to distant Van der Waals interactions and Josephson junctions between neighboring primary effectons. *This second order phase transition* is accompanied by (a)-states probability increasing ($P_a \to 1$) and that of (b)-states decreasing ($P_b \to 0$). The probability of primary and secondary deformons ($P_d = P_a \cdot P_b$; $\bar{P}_d = \bar{P}_a \cdot \bar{P}_b$) decreases correspondingly. In the excitation spectrum (Fig.1) these processes are displayed as a tending of (b)-branch closer to (a)-branch, as a consequence of degeneration of b-branch at very law temperature.

Like in the theory of 2nd order phase transitions proposed by Landau (Landau and Lifshits, 1976), we can introduce here the *order parameter* as:

$$\eta = 1 - \kappa = 1 - \frac{P_a - P_b}{P_a + P_b} \qquad\qquad 58$$

where: $\kappa = \frac{P_a - P_b}{P_a + P_b}$ is an equilibrium parameter.

One can see that at $P_a = P_b$, the equilibrium parameter $\kappa = 0$ and $\eta = 1$ (the system is far from 2nd order phase transition). On the other hand, at conditions of phase transition: $T \to T_\lambda$ when $P_b \to 0$, $\kappa \to 1$ and parameter of order tends to zero ($\eta \to 0$).

Similar to Landau's theory, the equality of specific parameter of order to zero, is a criteria of 2nd order phase transition. As usual, this transition is followed by a decrease in structural symmetry with a decline in temperature.

The important point of our scenario of superfluidity is a statement that the leftward shift of ($a \Leftrightarrow b$) equilibrium of the primary effectons (tr and lb) becomes stable starting from $T_\lambda$ due to their polymerization "side by side". This process of *macroscopic* Bose-condensation, including conversion of secondary effectons to primary ones, differs from condensation of an ideal Bose-gas described by eq. (1.26). Such kind of Bose-condensation means the enhancement of the concentration of primary effectons in (a) state with lower energy, accompanied by degeneration of the all other kind of collective excitations. The polymerization of primary effectons in He II gives rise to macroscopically long filament-like (or chain-like) polyeffectons.

*Such process can be considered as self-organization on macroscopic scale. These filament-like polyeffectons, representing superfluid component, can form closed circles or three-dimensional (3D) isotropic networks in a vessel with He II. The remnant fraction of liquid represent normal fraction of He II.*

Polyeffectons are characterized by the dynamic equilibrium: $\left[\, assembly \Leftrightarrow disassembly \,\right]$. Temperature decreasing and pressure increasing shift this equilibrium to the left, increasing the surface of the primary effectons side-by-side interaction and number of Josephson junctions. The probability of tunneling between coherent clusters increases also correspondingly.

The relative movement (sliding) of flexible "snake-like" polyeffectons occurs without phonons excitation in the volumes of IR deformons, equal to that of macrodeformons. Just macrodeformons excitation is responsible for dissipation and viscosity in normal liquids (Kaivarainen, 2001). The absence of macrodeformons excitation, making it possible the polyeffectons emergency (macroscopic Bose condensation), explains the superfluidity phenomenon according to our model.

Breaking of symmetry in a three-dimensional polyeffecton network and its violation can be induced by external fields, like the gravitational gradient, mechanical perturbation and surface effects. It is possible because coherent polyeffecton system is highly cooperative and unstable.

In rotating cylindrical vessel, filament-like polyeffectons originate from 3D isotropic net of polyeffectons and they tend to be oriented along the cylinder axis with their own rotation round their own axises in the direction opposite to that of cylinder rotation, as a consequence of angular momentum conservation. In accordance with our model, this phenomenon represents the vortex filaments in He II, discussed above. The radius of the filaments (42) is determined by the group velocity of the coherent $^4$He atoms, which form part of the primary effectons($\mathbf{v}_{gr} = \mathbf{v}_{sf}$). The numerical value of $\mathbf{v}_{gr}$ must be equal to or less than $6 \cdot 10^3 cm/s$, this corresponding to conditions (55 and 56). At $T \to 0$, $\mathbf{v}_{gr}$ decreases and the filament radius (42) increases to reach the values corresponding to $\mathbf{v}_{gr}^{\min} = \mathbf{v}^0$ determined by the zero-point oscillations of $^4$He atoms. Under these conditions the aggregation or polymerization of translational primary effectons in (a)-state can occur, leading to liquid-solid phase transition in $^4$He.

The self-organization of highly cooperative coherent polyeffectons in $\lambda$-point and strong

($\mathbf{a} \rightleftharpoons \mathbf{b}$) equilibrium leftward shift should be accompanied by a heat capacity jump.

The mechanism, leading to stabilization of (a)- state of primary effectons as the first stage of their polymerization, is a formation of *coherent superclusters* from primary effectons. Stabilization of (a) states in **superclusters** could be resulted from macroscopic self-organization of matter, turning mesoscopic Bose condensation to macroscopic one. Corresponding process stabilize the acoustic (a) state of primary effectons and destabilize the optic (b) state.

*The successive mechanisms of super-clusterization of primary effectons and polymerization of these superclusters could be responsible for second order phase transitions, leading to emergency of superfluidity and superconductivity.*

*The second sound* in such a model can be attributed to phase velocity in a system of polyeffectons or superclusters. The propagation of the second sound through chain polyeffectons or superclusters should be accompanied by their elastic deformation and [assembly ⇔ disassembly] equilibrium oscillations.

*The third sound* can be also related to the elastic deformation of polyeffectons and equilibrium constant oscillations of superclusters, *however only in the surface layers* with properties different from those in bulk volume. In accordance with hierarchic theory of surface tension for regular liquids (Kaivarainen, 2001), such a difference between surface and volume parameters is responsible for surface tension ($\sigma$) in quantum liquid, like **HeII**, and its increasing at $\lambda$-point. Such enhancement of $\sigma$ explains disappearance of cavitational bubbles at $T < T_\lambda$.

*The fourth sound* is a consequence of primary effectons volume increasing and the change in their phase velocity as a result of He II interaction with narrow capillary's walls and their thermal movement stabilization.

*The normal component of He II* is related to the fraction of He II atoms, not involved in polyeffectons formation. This fraction composes individual primary and secondary effectons, maintaining the ability for ($a \Leftrightarrow b$) and ($\bar{a} \Leftrightarrow \bar{b}$) transitions. In accordance with our hierarchic model, these transitions in composition of macroeffectons and macrodeformons are accompanied by the emission and absorption of heat phonons.

The manifestation of viscous properties in normal liquid and normal component of He II is related to fluctuations of macrodeformons ($V_D^M$), accompanied by dissipation (Kaivarainen, 2001).

On the other hand, macro- and superdeformons are absent in the superfluid component, as far in primary polyeffectons at $T < T_\lambda$: the probability of B-state of macroeffectons: $P_B = P_b \cdot \bar{P}_b \to 0$; the probability of A-state of the macroeffectons: $P_A = P_a \cdot \bar{P}_a \to 1$ and, consequently, the probability of macrodeformons tends to zero: $P_D^M = P_B \cdot P_A \to 0$. Decreasing the probability of superdeformons $P_D^S = (P_D^M)_{tr} \cdot (P_D^M)_{lb} \to 0$ means the decreased concentration of cavitational bubbles and vapor pressure.

**3.** We can explain the decrease in E(k) in Fig. 1 around $T = T_\lambda$ by reducing the contributions of (b) state of the primary effectons, due to their Bose-condensation, decreasing the fraction of secondary effectons and concomitant elimination of the contribution of secondary acoustic deformons (i.e. phonons) to the total energy of liquid $^4$He. One can see from our theory of viscosity (Kaivarainen, 1995; 2001), that in the absence of secondary effectons and macroeffectons excitations, providing dissipation in liquids, the viscosity of liquid tends to zero: $\eta \to 0$. In accordance with hierarchic theory of thermal conductivity (Kaivarainen, 1995; 2001), the elimination of secondary acoustic deformons at $T \leq T_\lambda$ must lead also to enhanced thermal conductivity. *This effect was registered experimentally in superfluids, indeed.*

**4.** The increase in $E(k)$ in Fig. 1 at $T < T_\lambda$ can be induced by the enhanced contribution of primary polyeffectons to the total energy of He II and the factor: $U_{tot}/T_k = S^{-1}$ in new state equation, derived in Hierarchic theory. The activity of the normal component of He II, as a solvent for polyeffectons, reduces and tends to zero at $T \to 0$. Under such condition ($T = 0$) super-polymerization and total Bose-condensation occur in the volume of $^4$**He**.

The maximum in Fig. 1 at $0 < T < T_\lambda$ is a result of competition of two opposite factors: rise in the total energy of **He II** due to progress of primary effectons polymerization and its reduction due to the decline of the most probable group velocity ($\mathbf{v}_{gr}$), accompanied by secondary effectons and deformons degeneration. The latter process predominates at $T \to 0$. The development of a polyeffectons superfluid subsystem is accompanied by corresponding diminution of the normal component in He II ($\rho_S \to 1$ and $\rho \to 0$). The normal component has a bigger internal energy than superfluid one.

The own dimensions of primary translational and librational effectons in composition of polyeffectons increases at $T \to 0$.

### 5.1 Inaccessibility of b-state of primary effectons at $T \leq T_\lambda$

Let us analyze our formula (Kaivarainen, 2001) for phase velocity of primary effectons in the acoustic (a)-state at condition $T \leq T_\lambda$, when filament - like polyeffectons from molecules of liquid originate:

$$\mathbf{v}_{ph}^a = \frac{\mathbf{v}_S \frac{1-f_d}{f_a}}{1 + \frac{P_b}{P_a}\left(\frac{v_{res}^b}{v_{res}^b}\right)} \qquad 59$$

where: $\mathbf{v}_S$ is the sound velocity; $P_b$ and $P_a$ are the thermoaccessibilities of the (b) and (a) states of primary effectons; $f_d$ and $f_a$ are the probabilities of primary deformons and primary effectons in (a) state excitations.

One can see from (59), that if:

$$P_b \to 0, \; then \; P_d = P_b P_a \to 0 \; and \; f_d \to 0 \; at \; T \leq T_\lambda$$

then phase velocity of the effecton in (a) state tends to sound velocity:

$$\mathbf{v}_{ph}^a \to \mathbf{v}_S \qquad 60$$

For these $\lambda -$ point conditions, the total energy of $^4\mathbf{He}$ atoms, forming polyeffectons due to Bose-condensation of secondary effectons (eq. 46), can be presented as:

$$E_{tot} \sim E_a = m\mathbf{v}_{gr}\mathbf{v}_{ph}^a \to m\mathbf{v}_{gr}\mathbf{v}_S \qquad 61$$

where the empirical sound velocity in He II is $\mathbf{v}_S = 2.4 \cdot 10^4 cm/s$.

The kinetic energy of wave B at the same conditions is $T_k = m\mathbf{v}_{gr}^2/2$. Dividing $E_{tot}$ by $T_k$ we have, using (47):

$$\frac{v_S}{v_{gr}} = \frac{E_{tot}}{2T_k} = \frac{1}{2S} = \frac{1}{2(m^*/m)} \qquad 62$$

and

$$\mathbf{v}_{gr}^0 = \mathbf{v}_s \cdot 2S^0 = 2.4 \cdot 10^4 \cdot 0.32 = 7.6 \cdot 10^3 cm/s. \qquad 63$$

$m^* = 0.16m$ is the semiempirical effective mass at $T = T_\lambda$.

The most probable wave B length corresponding to (63) at $\lambda$-point:

$$\lambda^0 = h/m\mathbf{v}_{gr}^0 = 15.1 \, \mathring{A} \qquad 64$$

The number of $^4\mathbf{He}$ atoms in the volume of such effecton, calculated in accordance with (54) is equal: $q^0 = (n_v^0)^{1/3} = 3.8$.

This result is even closer to one predicted by the hierarchic model (eq. 55) than (53). It confirms that at $T \leq T_\lambda$ the probability of b-state $P_b \to 0$ and conditions (60) and (61), following from our model, take a place indeed. In such a way our hierarchic model of superfluidity explains the available experimental data on liquid $^4$He in a non contradiction manner, as a limit case of our hierarchic viscosity theory for normal liquids.

## 5.2 *Superfluidity in $^3He$*

The scenario of superfluity, described above for Bose-liquid of $^4$**He** ($S = 0$) in principle is valid for Fermi-liquid of $^3$**He** ($S = \pm 1/2$) as well. A basic difference is determined by an additional preliminary stage related to the formation of Cooper pairs of $^3$He atoms with total spins, equal to 1, i.e. with boson's properties. The bosons only can form primary effectons, as a coherent clusters containing particles with *equal* energies.

We assume in our model that Cooper's pairs can be formed between neighboring $^3$**He** atoms. It means that the minimum number of $^3$**He** atoms forming part of the primary effecton's edge at $\lambda$-point must be 8, i.e. two times more than that in $^4$**He** (condition 55). Correspondingly, the number of $^3$**He** atoms in the volume of an effecton is $(n_F^V)_{3_{He}} = 8^3 = 312$. These conditions explains the fact that superfluidity in $^3$**He** arises at temperature $T = 2.6 \cdot 10^{-3}K$, i.e. much lower than that in $^4$**He**. The formation of flexible filament-like polyeffectons, representing macroscopic Bose-condensate in liquid $^3$**He** responsible for superfluidity, is a process, similar to that in $^4$**He** described above.

# 6. Superconductivity

## 6.1 *General properties of metals and semiconductors*

The dynamics of conductance electrons in metals and semiconductors is determined by three main factors (Kittel, 1978, Ashkroft and Mermin, 1976, Blakemore, 1985):

1. *The electric field influencing the energy of electrons;*
2. *The magnetic fields changing the direction of electrons motions;*
3. *Scattering on the other electrons, ions, phonons, defects.*

The latter factor determines the values of the electron conductance and resistance.

In spite of the small mean distances between electrons in metals (2-3)Å their mean free run length at room temperatures exceeds $10^4 \mathring{A}$ and grows by several orders at $T \to 0$. It is related to the fact that only electrons having energy higher than Fermi energy ($\epsilon_F$) may be involved in collisions. The fraction of these electrons in the total number of electrons is very small and decreases on lowering the temperature as $(kT/\epsilon_F)^2$. At room temperatures the scattering of electrons in metals occurs mainly on phonons. The mean free run length of electrons in indium at 2K is about 30 cm.

The analysis of electric and magnetic fields influence on the electrons needs the notion of its effective mass ($m^*$). It is introduced as a proportionality coefficient between the force acting on the electron and the acceleration (a) in the electric field (E):

$$F = -eE = m^*a; \qquad a = d\mathbf{v}_{gr}/dt \qquad\qquad 65$$

In a simple case of an isotropic solid body the effective mass of an electron is a scalar (Kittel, 1978):

$$m^* = \frac{\hbar^2}{d^2\epsilon/dk^2} \qquad\qquad 66$$

where $\epsilon$ is the kinetic energy of an electron, having a quadratic dependence on the wave number ($k = 1/L_B$):

$$\epsilon = \frac{\hbar^2 k^2}{m^*} = \frac{\hbar^2}{2mL^2} \qquad\qquad 67$$

In a general case, for electrons in solid bodies with a complex periodic structure, the effective mass is a tensor:

$$[m^*_{ij}] = \hbar^2/[\partial^2\epsilon/\partial k_i \partial k_j] \qquad\qquad 68$$

The effective mass tensor can have positive components for some directions and negative for others.

## 6.2  Plasma oscillations

At every displacement of the electron gas relative to subsystem of ions in solid body, the returning electric field appears. As a consequence of that, the subsystem of electrons will oscillate relative to the subsystem of ions with the characteristic plasma frequency (Ashkroft and Mermin, 1976):

$$\omega_{pl} = 2\pi v_{pl} = \left( \frac{4\pi ne^2}{m^*} \right)^{1/2} \qquad\qquad 69$$

where: (n) is the number of electrons in $1 cm^3$, (e) is the charge and ($m^*$) is the effective mass of an electron.

The quantified collective oscillations of electron plasma are termed *plasmons*. With decreasing (n) from $10^{22}$ to $10^{10} cm^{-3}$ the frequencies $\omega_{pl}$ decrease from $6 \cdot 10^{15} s^{-1}$ to $6 \cdot 10^3 s^{-1}$. For metals $\omega_{pl}$ corresponds to an ultraviolet frequency range, and for semiconductors - to an IR frequency range.

For longitudinal plasma oscillations at small wave vectors the dependence of frequency on the wave number ($k = 1/L = 2\pi/\lambda$) can be approximately represented as (Kittel, 1978):

$$\omega \approx \omega_{pl} \cdot \left( 1 + \frac{3k^2 \mathbf{v}_F^2}{10\omega_{pl}^2} + \dots \right) \qquad\qquad 70$$

where: $\mathbf{v}_F$ is the Fermi velocity of an electron (see eq.77).

The screening length ($l$), characterizing the electron-electron interaction in plasmon when Fermi-gas is degenerated, is equal to:

$$l = \mathbf{v}_F/v_p \sim 1\mathring{A} \quad \text{in metals} \qquad\qquad 71$$

For the cases of non-degenerated Fermi-gas, when the concentrations of free electrons are sufficiently low (in semiconductors) or at high temperatures $T \sim 10^4 K$, the screening length ($l_d$) is dependent on thermal electron velocity:

$$\mathbf{v}_{th} = \left( 3k_B T/m^* \right)^{1/2} \qquad\qquad 72$$

and

$$l_D = \mathbf{v}_{th}/v_{pl} \cong \left( \frac{\epsilon k_b T}{4\pi ne^2} \right)^{1/2} \qquad\qquad 73$$

where: $v_{pl}$ corresponds to (69), $\epsilon$ is the dielectric constant.

For example, if in a semiconductor $n = 5 \cdot 10^{17} cm^{-3}$ and $\epsilon = 12$, then $l_D = 60\mathring{A}$ (March, Parrinello, 1982).

## 6.3  Fermi energy

The notion of Fermi energy ($\epsilon_F$) can be derived from the Pauli principle forbidding the fermions to be in the same energetic states. The formula for Fermi energy for the case of ideal electron gas includes the electron mass (m), the Plank constant ($h = 2\pi\hbar$) and the concentration of free electrons ($n_e = N_e/V$):

$$\epsilon_F = \frac{h^2}{2m} \left( \frac{3}{8} \pi m_e \right)^{3/2} = \frac{2\pi^2\hbar^2}{m} \left( \frac{3}{8} \pi m_e \right)^{3/2} \qquad\qquad 73a$$

where $N_e$ is the number of free electrons in selected volume ($V$). For a real electron gas, $m$ must be substituted by its effective mass: $m \rightarrow m^*$.

The formula (73a) can also be derived using the idea of standing waves B of the unbind electrons of matter. The condition under which the concentration of twice polarized standing waves B of electrons is equal to the concentration of electrons themselves:

$$n_B^F = \frac{N_e}{V} = \frac{8\pi}{3(\lambda_B^F)^3} \qquad 74$$

The wave B length of an electron corresponding to this condition is:

$$\lambda_B^F = \left( \frac{8\pi V}{3N_e} \right)^{1/3} = \frac{h}{mv_{gr}^r} \qquad 75$$

The kinetic energy of the unbind electrons waves B ($T_k$) could be expressed through their length and mass. It appears that the kinetic energy of the electrons standing waves B, limited by their concentration is equal to Fermi energy:

$$T_k^F = \frac{h^2}{2m\lambda_F^2} = \frac{h^2}{2m} \left( \frac{3n_e}{8\pi} \right)^{2/3} = \frac{P_F^2}{2m} = \epsilon_F, \qquad 76$$

where Fermi momentum:

$$P_F = mv_F = h \left( \frac{3n_e}{8\pi} \right)^{1/3} = \hbar (3\pi^2 n_e)^{1/3} \qquad 77$$

The Fermi energy corresponds to Fermi temperature ($T_F$):

$$\epsilon_F = kT_F = hv_F \qquad 78$$

At $T < T_F$ electron gas is in a strongly "compressed" state. The more the relation $(T/T_F) = kT/\epsilon_F$, the more the probability of the appearance of "free volume" in a dense electron gas. On lowering the temperature, when the momentum of electrons decreases and the heat wave B length increases, the "effective pressure" of the electron gas grows, leading to its Bose-condensation.

### 6.4 *Cyclotronic resonance*

The magnetic field $B_z$ in the direction (z) influencing the electron due to Lorentz force, changes the direction of its motion without changing the energy. If electron's energy does not dissipate, then the electrons rotate in the plane xy, around z-axis. Such an electron with the effective mass $m^*$ has a circulation orbit of the radius $r$, with rotation frequency $\omega_c$. From the condition of equality between the Lorentz force ($r\omega_c eB_z$) and the centrifugal force ($m^*\omega_c^2 r$) the formula is derived for angular cyclotron frequency (Kittel, 1978):

$$\omega_c = eB_z/m^* \qquad 79$$

The kinetic energy, corresponding to the rotation is equal to:

$$T_k = \frac{1}{2} m^* (\omega_c)^2 r^2 \qquad 80$$

In the range of radio-frequencies ($\omega$) such a value of the magnetic induction $B_z$ can be selected that at this value the resonance energy absorption occurs, when $\omega = \omega_c$.

Such experiments on the cyclotron resonance can be done to determine $m^*$ in selected directions.

In a simple case, an electron revolves around the Fermi sphere with the *zero momentum component in z-direction.* The radius of this sphere is determined by the Fermi momentum $P_F$ (see eq. 77). In the real space:

$$r_F \sim \hbar/P_F \qquad 81$$

The energy of free particles near the Fermi surface:

$$\epsilon(P_F) = \mathbf{v}_F(P - P_F)$$   82

where: $\mathbf{v}_F$ and $P_F = m^*\mathbf{v}_F$ are the Fermi velocity and momentum: $P > P_F$ is the momentum of thermal electron at $T > 0$ near the Fermi surface.

The solution of the Schrödinger equation, modified by Landau for electrons in a magnetic field in real space, leads to following total energy eigenvalues (Blakemore, 1985):

$$\epsilon = \frac{\hbar^2 k_z^2}{m^*} + \left(l + \frac{1}{2}\right)\hbar\omega_c,$$   83

where the first term of right part represent the energy of *translational* motion of electrons, which does not depend on magnetic field magnitude; $k_z = 1/L_z$ is the wave number of this motion; the second term is responsible for rotational energy, $l = 0, 1, 2\ldots$ is the integer quantum number for rotational motion in magnetic field $B_z$. Every value of $l$ means a corresponding Landau level.

Thus, free electrons in a magnetic field move along the helical trajectory of the radius:

$$r_l = [(2l+1)\hbar/m^*\omega_c]^{1/2}$$   84

At the transition from real space to the wave number space, the radius of the orbit ($k_p$) and its area is quantified as:

$$S_l = \pi k_p^2 = \frac{2\pi e B}{\hbar c}\left(l + \frac{1}{2}\right)$$   85

This formula is valid not only for the free electron model, but also for real metals. The magnitude $2\pi(\hbar c/e)$ termed a flux quantum. In a strong magnetic field the quantization of electrons energy leads to the periodic dependence of the metal's magnetic moment on magnetic field tension (B): the de Haaz - van Alfen effect (Kittel, 1978, Ashkroft and Mermin, 1976).

### 6.5 Electroconductivity

According to the Sommerfeld theory (Blakemore, 1985), electroconductivity ($\sigma$) depends on the free run time of an electron ($\tau$) between collisions:

$$\sigma = ne^2\tau/m,$$   86

where: n is the concentration of electrons, (e) and (m) are electron charge and mass.

The free run time is equal to the ratio of the average free run distance ($\lambda$) of electrons to the Fermi speed ($\mathbf{v}_F$):

$$\tau = \lambda/\mathbf{v}_F$$   87

The free run distance is determined by scattering at defects ($\lambda_D$) and scattering at phonons ($\lambda_{ph}$):

$$1/\lambda = 1/\lambda_D + 1/\lambda_{ph}$$   88

The resistance ($R = 1/\sigma$) could be expressed as:

$$R = 1/\sigma_D + 1/\sigma_{ph} = R_D + R_{ph}$$   89

the contribution $R_D$ depends mainly on the concentration of the conductors defects, and the phonon contribution $R_{ph}$ depends on temperature. Formula (89) expresses the Mattisen rule.

*The transition to superconducting state* means that the free run time and distance tend to infinity: $\tau \to \infty$; $\lambda \simeq \lambda_{ef} = h/P_{ef} \to \infty$, while the *resulting* group velocity of the electrons ($\mathbf{v}_{gr}^{res}$) and momentums tends to zero:

$$P_{ef} = m \cdot \mathbf{v}_{gr}^{res} \to 0$$   89a

The emergency of macroscopic Bose-condensation of ionic effectons and electron's Cooper pairs, accompanied by slowing down the secondary electronic effectons with nonzero momentum to primary ones, corresponds to condition (89a). We assume in our model, that in this case, the absence of the non-elastic scattering and dissipation of electrons energy is resulted in superconductivity.

*Let us consider at first the conventional microscopic approach to problem of superconductivity.*

## 7. Microscopic theory of superconductivity (BCS)

This theory (BCS) was created by Bardin, Cooper and Schriffer in 1957. The basic, experimentally proven assumption of this theory, is that electrons at sufficiently low temperatures are grouped into Cooper pairs with oppositely directed spins - Bose-particles with a zero spin. The charge of the pair is equal to $e^* = 2e$ and mass $m^* = 2m_e$.

Such electron's pairs obey the Bose-Einstein statistic. The Bose- condensation of this system at the temperature below the Bose-gas condensation temperature ($T < T_k$) leads to the superfluidity of the electron liquid. This superfluidity(analogous to the superfluidity of liquid helium) is manifested as superconductivity.

According to BCS's theory, the Cooper electron pair formation mechanism is the consequence of virtual phonon exchange through the lattice.

The energy of binding between the electrons in a pair is very low: $2\Delta \sim 3kT_c$. It determines a minimum energetic gap ($\Delta$), separating state of superconductivity from state of regular conductivity.

Notwithstanding that kinetic energy of the electrons in a superconducting state is greater than $\epsilon_F$, the contribution of the potential energy of attraction between electron pairs is such, that the total energy of the superconducting state ($E^e$) is smaller than the Fermi energy ($\epsilon_F$) (Kittel, 1978). The corresponding energetic gap $\Delta$ makes superconducting state stable after switching-off external voltage. The middle of the gap coincides with the Fermi level.

The rupture of a pair can happen due to photon absorption by superconductor with the energy: $\Delta = \epsilon_F - E^e = h\nu_p \approx 3kT_c$. Superconductivity usually disappears in the frequency range $10^9 < \nu_p < 10^{14} s^{-1}$.

In the BCS theory, the magnitude of $\Delta$ is proportional to number of Cooper pairs and grows up on lowering the temperature.

The excitation energy of quasiparticles in a superconducting state, which is characterized by the wave number (k), is:

$$E_k = (\epsilon_k^2 + \Delta^2)^{1/2},$$ 90

where

$$\epsilon_k = \frac{\hbar^2}{m}(k^2 - k_F^2) \approx \frac{\hbar^2}{m} k_F(k - k_F)$$ 91

and

$$\delta k_F = (k - k_F) \ll k_F = 1/L_F$$

The critical speed of the electron gas ($\mathbf{v}_c$), for exciting a transition from a superconducting state to a normal one is determined from the condition:

$$E_k = \hbar k \mathbf{v}_c \text{ and } \mathbf{v}_c = \frac{E_k}{\hbar k}$$ 92

The wave function $\Phi(\mathbf{r})$, which describes the properties of electron pairs in the BCS theory, is a superposition of one-electron functions with energies in a range of about $2\Delta$ near $\epsilon_F$. Therefore, the dispersion of momentums for one-electron levels, involved in the formation of pairs, is expressed as:

$$\Delta = \delta\epsilon_F = \delta\left(\frac{P_F^2}{2m}\right) = \left(\frac{P_F}{m}\right)\delta P_F \approx v_F \delta P_F \qquad 93$$

where: $P_F$ and $\mathbf{v}_F$ are the Fermi momentum and velocity.

The characteristic coherence length ($\xi_c$) of the pair function $\Phi(r)$ has a value (Ashkroft and Mermin, 1976, Lifshits and Pitaevsky, 1978):

$$\xi_c \ \sim \hbar/\delta P_F \simeq \frac{\hbar v_F}{\Delta} \ \simeq \ \frac{1}{k_F}\frac{\epsilon_F}{\Delta} \qquad 94$$

The magnitude ($\epsilon_F/\Delta$) is usually $10^3 - 10^4$, and $k_F = 1/L_F \sim 10^8 cm^{-1}$. Thus, from (94):

$$\xi_c \sim (10^3 - 10^4)\AA \qquad 95$$

Inside the region of coherence length ($\xi_c$) there are millions of pairs. The momentums of pairs in such regions are correlated in such a way that their resulting momentum is equal to zero.

At T > 0 some of the pairs turn to a dissociated state and the concentration of superconducting electrons ($n_s$) decreases. The coherence length ($\xi_c$) also tends to zero with increase in temperature.

The important parameter, characterizing the properties of a superconductor, is the value of the critical magnetic field ($H_c$), above which the superconductor switches to a normal state.

With a rise in the temperature of the superconductor when $T \to T_c$, the critical field tends at zero: $H_c \to 0$. And vice versa, at lowering of temperature, when

$T < T_c$, the $H_c$ grows up as:

$H_c = H_0[1 - (T/T_c)^2]$

where $H_0$ corresponds to $T = 0$.

The Meisner effect - the "forcing" of the outer magnetic field out of the superconductor, is also an important feature of superconductivity.

The depth of magnetic field penetration into the superconductor ($\lambda$) (Kittel, 1978, Ashkroft and Mermin, 1976) is:

$$\lambda = (m\epsilon_0 c^2/e^2 n_s)^{1/2} \simeq (10^{-6} - 10^{-5})cm,$$

where $n_s$ the density of electrons in a superfluid state; $\epsilon_0$ - dielectric constant.

On temperature raising from 0 K to $T_c$, the $\lambda$ grows as:

$$\lambda = \frac{\lambda_0}{\left[1 - \left(\frac{T}{T_c}\right)^4\right]^{1/2}}, \qquad 96$$

where: $\lambda_0$ corresponds to $\lambda$ at $T = 0$.

The superconductors with magnetic field penetration depth ($\lambda$), less than coherence length $\xi$:

$$\lambda \ll \xi \qquad 97$$

are termed first order superconductors and those with

$$\lambda \gg \xi \qquad 98$$

are termed second order superconductors.

Nowadays, in connection with the discovery of high temperature superconductivity (Bednorz, Muller, 1986, Nelson, 1987) the mechanism of stabilizing electron pairs by means of virtual phonons in the BCS theory evokes doubts.

## 8. Hierarchic scenario of superconductivity

We propose a new mechanism of electron's Cooper pairs macroscopic Bose-condensation, without virtual phonons as mediators. Such a process is analogous to formation of primary polyeffectons in liquid helium, in our explanation of superfluidity (see Section 5).

Two basic questions must be answered in relation to appearance of superconductivity:

I. Why does energy dissipation in the system [conductivity electrons + lattice] disappear at $T \leq T_c$ ?

II. How does the coherence in this system, related to electron pair formation, originate ?

*It will be shown below how these problems can be solved in the framework of our Hierarchic theory of matter (Kaivarainen, 2001; 2001a; 2003).*

Following factors can affect the electron's dynamics and scattering near Fermi energy:

1. Interaction of electrons with primary and secondary ionic effectons of lattice in acoustic (a) and (ā) states, stimulating Cooper pairs formation;

2. Interaction of electrons with primary and secondary effectons of lattice in optic (b) and (b̄) states, inducing Cooper pairs dissociation;

3. Interaction with ($a \Leftrightarrow b$) and ($\bar{a} \Leftrightarrow \bar{b}$) transitons of primary and secondary effectons;

4. Interaction with [tr/lb] convertons (interconversions between primary translational and librational effectons;

5. Interaction with primary electromagnetic deformons;

6. Interaction with secondary acoustic deformons (possibility of polaron formation);

7. Interaction with macroeffectons in A- and B-states;

8. Interaction with macro- and superdeformons (possible appearance of defectons).

It follows from our model that the oscillations of all types of quasiparticles in conductors and semiconductors are accompanied not only by electron-phonon scattering, but also by electromagnetic interaction of primary deformons with unbind electrons.

At $T > T_c$ the fluctuations of unbind electrons with energy higher than Fermi one under the influence of factors (1 - 8) are random (noise-like) and no selected order of fluctuations in normal conductors exists. It means that ideal Fermi-gas approximation for such electrons is sufficiently good. In this case, the effective electron mass can be close to that of a free electron ($m^* \simeq m$).

Electric current in normal conductors at external voltage should dissipate due to fluctuations and energy exchange of the electrons with lattice determined by factors (1 - 8).

Coherent in-phase acoustic oscillations of the ionic primary effectons in (a)-state is the "ordering factor" simulating electron gas coherence due to long-range Van-der-Waals and electromagnetic interactions. But their contribution in regular conductors at $T > T_c$ is very small. For ideal electron gas, the total energy ($E_{tot}$) of each electron (de Broglie wave) is equal to its kinetic energy ($T_k$), as far potential energy ($V = 0$):

$$E_{tot} = \hbar \omega = \frac{\hbar^2}{2mA^2} = T_k + V = \frac{\hbar^2}{2mL^2} = T_k \qquad 99$$

where:

$$T_k = \frac{\hbar^2}{2mL^2} \qquad 100$$

and

$$L = \frac{\hbar}{mv_{gr}} = \frac{1}{k} \qquad 101$$

One can see from (100), that for an ideal gas, when $m = m^*$, the most probable amplitude (A) and length (L) of de Broglie wave (wave B) are equal:

$$A = L, \text{ if } E_{\text{tot}} = T_k \text{ and } V = 0 \qquad 102$$

Like in case of liquid helium at conditions of superfluity, Bose- condensation in metals and semiconductors is related to increasing of concentration of the (a)-state of primary effectons with the lowest energy and a corresponding decreasing of concentrations of all other excitations. The Bose-condensation and degeneration of secondary ionic effectons and deformons, followed by formation of electronic polyeffectons from Cooper pairs, is responsible for second order phase transition, like superconductivity.

The cooperative character of 2nd order phase transition [conductor → superconductor] is determined by a feedback reaction between the lattice and electron subsystems. It means that the collective *macroscopic* Bose-condensation from *mesoscopic* Bose condensation (mBC) in both subsystems is starting at the same temperature: $T = T_c$.

Under such conditions the probabilities of the (a)-states of ionic ($P_a^i$) and electronic ($P_a^e$) primary effectons tend to 1 at $T \leq T_c$:

$$\left. \begin{array}{l} P_a^i \to 1; \ P_a^e \to 1 \\ P_b^i \to 0; \ P_b^e \to 0 \end{array} \right\} \qquad 103$$

The equilibrium parameter for both subsystems:

$$\kappa^{i,e} = \left( \frac{P_a - P_b}{P_a + P_b} \right)^{i,e} \to 1 \qquad 104$$

and the *order parameter*:

$$\eta^{i,e} = (1 - \kappa^{i,e}) \to 0 \qquad 105$$

like in a *2nd* order phase transition (superfluidity) for liquid helium (see eq. 58).

*In our model the coherent Cooper pairs with resulting spin, equal to 0 and 1, are formed from two neighboring electrons, in contrast to the BCS theory, which assumes phonons - mediated interaction between two distant electrons.*

In turn, the electron's Cooper pairs can compose primary electronic effectons (e-effectons), as a coherent cluster with external momentum equal to zero, with linear dimension, equal to dimension of Cooper pair in BCS theory or coherence length. The interaction of e-effectons, having nonzero external momentum, with lattice is responsible for electric resistance in regular conductors. Degeneration of such type of excitations in the process of their Bose-condensation and their conversion to primary e-effectons means the emergency of superconductivity.

The in-phase coherent oscillations of the integer number of electron pairs, forming primary e-effectons, correspond to its acoustic (a)- state, and the counterphase oscillations - to its optic (b) state.

We assume that superconductivity can originate only when the fraction of unbind coherent electrons, forming primary e-effectons and polyeffectons in certain regions of conductor, strongly prevails over the fraction of noncoherent secondary electron's effectons. Due to feedback reaction between subsystems of lattice and electrons this fraction should be equal to *ratio of wave length of primary and secondary ionic effectons*. This condition can be introduced as:

$$(II) : \left( \frac{\lambda_a^e}{\bar{\lambda}_a^e} \right)_{T_c} = \left( \frac{\lambda_a^i}{\bar{\lambda}_a^i} \right)_{T_c} = \left( \frac{v_s/\lambda_a}{v_s/\bar{\lambda}_a} \right)_{T_c} \geq 10 \qquad 106$$

where: $\lambda_a^e / \bar{\lambda}_a^e$ is the ratio of wave lengths of primary and secondary e- effectons, equal to that of primary and secondary effectons of lattice; sound velocity is equal to lattice (ionic) primary effectons phase velocity in (a) state: $v_s \simeq v_{ph}^a$ under conditions corresponding to (60).

As far the oscillations of e-pairs in Bose-condensate (a-state of e- effectons) are modulated by the electromagnetic field, radiated by oscillating ions in the (a)-state of primary ionic effectons, they must have the same frequency.

Condition (106) means that the number of electrons in the volume ($V_e \sim \lambda^3$) of primary e-effectons is about $10^3$ times more than that in secondary e-effectons. The lattice and electronic effectons subsystems are spatially compatible.

As far the effective mass ($m^*$) of the electrons in macroscopic Bose-condensate, formed by electronic polyeffectons, tends to infinity:

$$
\begin{aligned}
& T < T_c \\
& m^* \mapsto \infty \\
& T \to 0
\end{aligned}
$$

108

the plasma frequency (eq. 69) tends to zero:

$$
\begin{aligned}
& T < T_c \\
2\pi v_{pl} = \omega_{pl} & \mapsto 0 \\
& at \; T \to 0
\end{aligned}
$$

109

and, consequently, the screening length ($eq.\,71$) tends to infinity: $l \to \infty$. This condition also corresponds to that of macroscopic Bose-condensation origination.

Under these conditions, the decrease of potential and total energy of both electron's and ion's subsystems due to leftward ($a \Leftrightarrow b$) equilibrium shift, leads to the appearance of the gap near the Fermi surface ($2\Delta$), depending on the difference of energy between (a) and (b) states of primary e-effectons.

The linear dimension of coherent primary electronic effectons (e-effectons), which is equal to coherence length in the BCS theory (see eq. 94), is determined by variation of Fermi momentum ($\Delta p_F$) :

$$
\xi = \lambda_a^e = \frac{\mathbf{v}_s}{V_a} = h/\Delta p_F = h/m_e \Delta \mathbf{v}_F
$$

110

In turn, the primary e-effecton in (a)-state can form e- polyeffectons as a result of their polymerization. The starting point of this collective process represents macroscopic Bose - condensation and second order phase transition, in accordance with our model.

The energy gap between normal and superconductive states can be calculated directly from our hierarchic theory, as the difference between the total energy of matter before $\left[\; U_{\text{tot}}^{T>T_c} \;\right]$ and after $\left[\; U_{\text{tot}}^{T<T_c} \;\right]$ the second order phase transition:

$$
2\Delta = U^{T>T_c} - U^{T<T_c}
$$

111

Such experimental parameters as *sound velocity, density and the positions of librational bands of IR spectrum* should be known around transition temperature ($T_c$) for calculation of (111), using Hierarchic theory based computer program (Kaivarainen, 2001).

### 8.1 *Interpretation of experimental data in the framework of our model of superconductivity*

The gap (111) must be close to the energy of $(a \to b)^i$ transitions of ionic primary effectons, correlated with energy of $(a \to b)^e$ transitions of the electronic e-effectons. This statement of our superconductivity model coincides well with known experimental fact: the violation of superconductivity state by IR-radiation occur at the minimum photons frequency ($v_p$), corresponding to energy gap value ($2\Delta$) at given temperature:

$$
hv_p = 2\Delta \sim (E_b - E_a)
$$

112

Another general feature of superconductivity for low- and high-temperature superconductors is almost constant ratio of energy gap to thermal energy at transition temperature ($T_c$):

$$\frac{2\Delta_0}{k_B T_c} \simeq 3.5 \qquad\qquad 113$$

where the minimum gap: $\Delta = \Delta_0$ at $T = T_c$ and $\Delta = 0$ at $T > T_c$.

It will be shown below, that the experimental result (113) is related to condition (106) of our hierarchic model of superconductivity.

Considering (112), (113) and (2.27), the frequency of a primary ionic effectons in a-state near transition temperature is:

$$\nu_a^i = \frac{\nu_g^0}{\exp\left(\frac{2\Delta}{kT_c}\right) - 1} \simeq \nu_g^0 / 32.1 = 0.03(2\Delta/h) \qquad\qquad 114$$

consequently:

$$h\nu_a^i = 0.03 \cdot 2\Delta \qquad\qquad 115$$

The frequency of *secondary* lattice effectons in $(\overline{a})$−state is (Kaivarainen, 2001):

$$\overline{\nu_a^i} = \frac{\nu_a^i}{\exp\left(\frac{h\nu_a^i}{kT_c}\right) - 1} \qquad\qquad 115a$$

as far:

$$\frac{h\nu_a}{kT_c} = 0.03 \frac{2\Delta}{kT_c} = 0.1 \ll 1 \qquad\qquad 116$$

we have: $\left[\exp\left(\frac{h\nu_a^i}{kT_c}\right) - 1\right]^{-1} \sim \frac{kT_c}{h\nu_a} \sim 10$

the frequency of *primary* lattice (ionic) effectons is:

$$\nu_a \simeq kT_c/h, \qquad\qquad 117$$

Now, using (17), (116) and (113) we confirm the correctness of introduced condition (106):

$$\lambda_a^e / \overline{\lambda_a^e} = (\overline{\nu}_a / \nu_a)^i = \left[\exp\left(\frac{h\nu_a^i}{kT_c}\right) - 1\right]^{-1} \sim 10 \qquad\qquad 118$$

where:

$$\lambda_a^e = h/2m_e(\mathbf{v}_{gr}^a)^e = (\mathbf{v}_s / \nu_a)^i \qquad\qquad 119$$

is the most probable wave B length of coherent electron's Cooper pairs, which determines the edge length of primary e-effecton; $(\mathbf{v}_{gr}^a)^e$ is a group velocity of electron pairs in a-state of primary e-effectons, oscillating in-phase with ionic lattice (ionic effectons in a-state) oscillations.

The mean wave B length of electron's pairs, forming the effective secondary e-effecton is:

$$\overline{\lambda_a^e} = h/2m^*\overline{\mathbf{v}_{gr}^a} = (\mathbf{v}_s / \overline{\nu_a})^i \qquad\qquad 120$$

*Our theory predicts also the 2nd condition of coherency between ionic and electronic subsystems, leading to superconductivity. This 'tuning' occur, when the linear dimension of primary translational ionic effectons grows up to the value of coherence length, equal in*

*accordance to our model, to the edge length of primary e-effecton (see eq.94):*

$$(\lambda_a^i)_{tr} = \frac{\mathbf{v}_s}{(v_a^i)_{tr}} \geq \frac{\zeta}{T_c} = \frac{\hbar v_F}{\Delta} \qquad 121$$

In simple metals a relation between sound ($\mathbf{v}_s$) and Fermi velocities ($\mathbf{v}_F$) is determined by the electron to ion mass ratio ($m_e/M$)$^{1/2}$(March and Parinello, 1982):

$$\mathbf{v}_s = \left( \frac{zm_e}{2M} \right)^{1/2} \cdot \mathbf{v}_F, \qquad 122$$

where: z is the valence of ions in a metal.

Putting $(v_a^i)_{tr}$ from eq.(115) and eq.(122) in condition (121), and introducing instead electron mass its effective mass ($m^*$) in composition of e-effecton , we obtain at $T = T_c$:

$$m^* \overset{T_c}{\simeq} 2 \cdot 10^{-4} \cdot \frac{M}{z} \qquad 123$$

As far we assume here that at transition temperature ($T_c$) the volumes of primary lattice (ionic) effectons and primary e-effectons coincide, then the number of electrons in the volume of primary effectons is:

$$N_e = n_e \cdot V_{ef}^i = n_e \cdot \frac{9}{4\pi} \left( \frac{\mathbf{v}_s}{v_a^i} \right)^3_{T_c} \qquad 124$$

where: $n_e$ is a concentration of the electrons; $V_{ef}^i = V_{ef}^e$ is the volume of primary translational ionic effectons at transition temperature.

Using the relation between primary and secondary ionic waves B (106) at $\mathbf{T}_c$ as $\lambda_a = 10\overline{\lambda}_a$, taking into account (117) and (121) we got:

$$\left( \frac{zm_e^*}{2} \right)^{1/2} \cdot \frac{1}{T_c M^{1/2}} = \frac{k}{10 \cdot 2\pi\Delta} \qquad 125$$

or:

$$T_c(M/z)^{1/2} = \frac{10\pi\Delta}{k} \cdot (2m_e^*)^{1/2} = const \qquad 126$$

where the transition temperature:

$$T_c \simeq \left( \frac{\hbar}{k_B} \right) \cdot \frac{\mathbf{v}_s}{2n_e^{1/3}} \qquad 127$$

As far, for different isotopes the energetic gap is constant ($2\Delta \simeq const$), then the left part of (126) is constant also. Such an important correlation between transition temperature ($T_c$) and isotope mass ($M$) is experimentally confirmed for many metals.

*This result, as well as (117), can be considered as evidence in proof for our model of superconductivity.*

It leads from (127) that the more rigid is lattice and the bigger is sound velocity ($\mathbf{v}_s$), the higher is transition temperature. Anisotropy of $\mathbf{v}_s$ means the anisotropy of superconductor properties and can be affected by external factors, such as pressure.

*We may conclude, that all most important phenomena, related to turbulence, superfluidity and superconductivity, can be explained in the framework of our Hierarchic theory of condensed matter, general for liquids and solids. This points once more to universal general properties of theory, making it possible the quantitative investigation of these complex phenomena, using new optoacoustic device: Comprehensive Analyzer of Matter Properties (CAMP), based on Hierarchic theory ( http://arxiv.org/abs/physics/0102086).*

# REFERENCES

Ashkroft N., Mermin N. Solid state physics. N.Y.:Helt, Rinehart and Winston, 1976.

Bardeen J., Cooper L.N., Schrieffer J.R. Phys. Rev., 108, 1175, 1957

Bardeen J., Schrieffer J.R. Progr. Low. Temp. Phys, 3, 170, 1960.

Beizer A. Basic ideas of modern physics. Nauka, Moscow, 1973.

Bednorz J.G., Muller K.A. Z.Phys.B. Condensed Matter, 64, 189, 1986

Blakemore J.S. Solid state physics. Cambrige University Press, Cambrige, N.Y. e.a, 1985.

Bogolyubov N.N. Lectures on quantum statistics. Collected works. Vol.2. Kiev, 1970.

Cooper L.N. Phys. Rev., 104, 1189.

Feynman R. Statistical mechanics.

Feynman R. The character of physical law. Cox and Wyman Ltd., London, 1965.

Frolich H. Phys. Rev., 79, 845, 1950.

Käiväräinen A.I. Theory of condensed state as a hierarchical system of quasiparticles formed by phonons and three-dimensional de Broglie waves of molecules. Application of theory to thermodynamics of water and ice. J.Mol.Liq. 1989$a$, 41, 53 – 60.

Kaivarainen A. Book: Hierarchic Concept of Matter and Field. Water, biosystems and elementary particles. New York, NY, 1995,  ISBN  0-9642557-0-7.

Käiväräinen A. New Hierarchic Theory of Condensed Matter and its Computerized Application to Water and Ice. Archievs of Los-Alamos: http://arxiv.org/abs/physics/0102086 (2001).

Kaivarainen A., Hierarchic theory of matter, general for liquids and solids: ice, water and phase transitions. American Institute of Physics Conference Proceedings, **573,** 181-200, (2001a).

Kaivarainen A.  New Hierarchic theory of water and its role in biosystems. The quantum Psi problem. Proceedings of the International conference: "Energy and informational transfer in biological systems: How physics could enrich biological understanding". F. Musumeci, L.S. Brishik, M.W. Ho (editors), World Scientific, ISBN 981-238-419-7, pp. 82-147, (2003a).

Kittel Ch. Thermal physics. John Wiley and Sons,Inc., N.Y., 1975.

Kittel Ch. Introduction to the solid state physics. Nauka, Moscow, 1978 (in Russian).

London F. On the Bose-Einstein condensation. Phys.Rev. 1938, 54, 947.

London F. Superfluids, v1, Wiley, 1950

Landau L.D., Lifshits E.M. Statistical physics. Nauka, Moscow, 1976 (in Russian).

Prokhorov A.M. (Ed.)Physical encyclopedia. Vol.1-4. Moscow, 1988.

Schrieffer J. 1957

Zeldovitch Ya.B., Khlopov M.Yu. Drama of concepts in cognition of nature. Nauka, Moscow, 1988.

See also:  http://arxiv.org/find/physics/1/au:+Kaivarainen/0/1/0/all/0/1